\documentclass[%
 reprint,
 amsmath,amssymb,
 aps,
]{revtex4-2}

\usepackage{graphicx}
\usepackage{dcolumn}
\usepackage{bm}
\usepackage{xcolor}

\begin{document}

\preprint{APS/123-QED}

\title{Unstable growth of bubbles from a constriction}

\author{Marc Grosjean}
\author{Elise Lorenceau}%
 
\affiliation{%
 Univ. Grenoble Alpes, CNRS, LIPhy, F-38000 Grenoble, France \\
}%

\date{\today}

\begin{abstract}
Bubbles and droplets are ubiquitous in many areas of engineering, including microfluidics where they can serve as microreactors for screening of chemical reactions.  They are often formed out of a constriction (a microfluidic channel or a cylindrical tube) by blowing a given volume of gas into a liquid phase. It is obviously crucial to be able to control their size, which is not always easy due to the coupling between the volume of the bubble and the gas pressure induced by the Laplace law.  In this paper, we examine the size and formation dynamics of soap bubbles blown from a cylindrical tube, which is the paradigm geometry for bubble and droplet formation. To do so, one end of the tube is closed by a soap film, while the other end is connected to a large reservoir of variable volume filled with gas. To inflate the gas in the bubble, we reduce the volume of the reservoir, which  mimics air inflation through the lung diaphragm or the flow-rate driven bubble formation in microfluidics geometry such as flow-focusing. As the volume of the reservoir decreases, the gas pressure increases, the soap film curves and takes the form of a spherical cap with an increasingly smaller radius of curvature. This quasi-static process continues until a critical pressure is reached for which the bubble is quasi-hemispherical. Beyond this pressure, the film undergoes a rapid topological transformation and swells very rapidly (in less than a hundred ms) until it reaches its final volume. We describe this instability in particular by showing that this unstable regime appears when a dimensionless number - whose expression we specify - reaches a critical value. Using a quasi-static model that we solve analytically, we predict the bubble growth dynamics and the final height of the bubble produced for any reservoir volume and constriction size.

\end{abstract}

\maketitle

\section{Introduction}

Among the successes of microfluidics is the possibility of forming large assemblies of drops or bubbles almost identical at high throughput (of the order of 100 Hz) \cite{Anna2003,Huerre2014,Anna2016}. These entities, dispersed in a continuous liquid phase and used as microreactors containing active ingredients at a concentration changing from drop to drop, allows analysis and screening of chemical reactions with unprecedented throughput \cite{Song2006,Brouzes2009,Guo2012,Baret2012}. Bubbles and drops are also found in other fields of engineering where they are generally dispersed in a continuous liquid phase, themselves then being qualified as the dispersed phase (fire-fighting foam or sparkling drinks \cite{Stevenson-foamengineering}). There are many methods to make monodisperse bubbles or drops such as shearing crude emulsion to split it into droplets\cite{Mason1996,Bibette2002} or blowing on an interface  \cite{Basaran2002,Salkin2016,Hamlett2021}. One commonly used in microfluidics and called flow-focusing consists in forming bubbles (or drops) by deforming an air/liquid interface placed at the end of a tube (of square, rectangular or circular section) from a reservoir whose pressure increases \cite{Anna2003}. In the dripping regime, three distinct steps can be identified: (1) a phase of quasi-static deformation of the interface fixed to the end of the tube which, by bending, changes from a flat geometry to that of a quasi-hemispherical cap with a radius equal to that of the tube. (2) a rapid growth (generally in less than 100 milliseconds) of this hemisphere until reaching a final almost spherical shape of radius much greater than that of the tube. (3) Finally, pinch-off regime with detachment of the bubble from the constriction \cite{Garstecki2005,Dollet2008}. Depending on the geometry of the system used, gravity or viscous friction forces produce the work necessary to stretch the neck separating the spherical bubble to the point of spontaneous rupture driven by capillary forces\cite{Eggers1997,Eggers2008}. When nothing disturbs the bubble, it remains attached to the end of the constriction or tube, as beautifully illustrated in the magnificent paintings "Les bulles de savon" of J.S. Chardin, E. Manet or J. Bail \cite{Les_bulles_de_savon}.

To obtain the most peaked bubble size distribution, the time of the pinch-off regime (3), which is intrinsically variable as a result of hydrodynamic instability, must be much shorter than the time of growth regimes (1) and (2). Thus, the pinching dynamics of fluid necks have been studied with great care, revealing the importance of convection \cite{ganan2004,Garstecki2005}, swirl \cite{Herrada2011}, confinement \cite{Hagedorn2004}, presence of surfactants or not \cite{vanHoeve2011}. On the contrary, the dynamics of phases (1) and (2) have been much less explored, the implicit hypothesis being that the duration of this phase is controlled by the flow rate of the dispersed phase and the volume of the bubble at the threshold of breakup. However, what sets this flow rate is not always obvious. For pressure-driven flow of the dispersed phase, nonlinear variations of the gas flow rate, induced by hydrodynamic feedback in the outlet channel have been reported in several studies \cite{Raven2006,Raven2006B,Sullivan2008}. For flow-rate driven flow, this difficulty should not exist, yet we reveal in this work that the compliance of the system - which arises here from the gas compressibility - induces a mechanical coupling between the deformation of the interface at the constriction and the pressure in the gas reservoir. This coupling can induce large fluctuations in the flow rate that lead to unstable bubble formation modes. We therefore study the first steps of bubble growth at an imposed flow rate in a geometry reminiscent of the one used by children when they blow a bubble from a tube. By comparing experimental results with an analytical model, we predict the final bubble volume. In particular, we show that the initial volume of the reservoir comprising the gas - usually not considered - is a key parameter of this process. 

\section{Experiment}

\begin{figure}
    \centering
    \includegraphics[width=9cm]{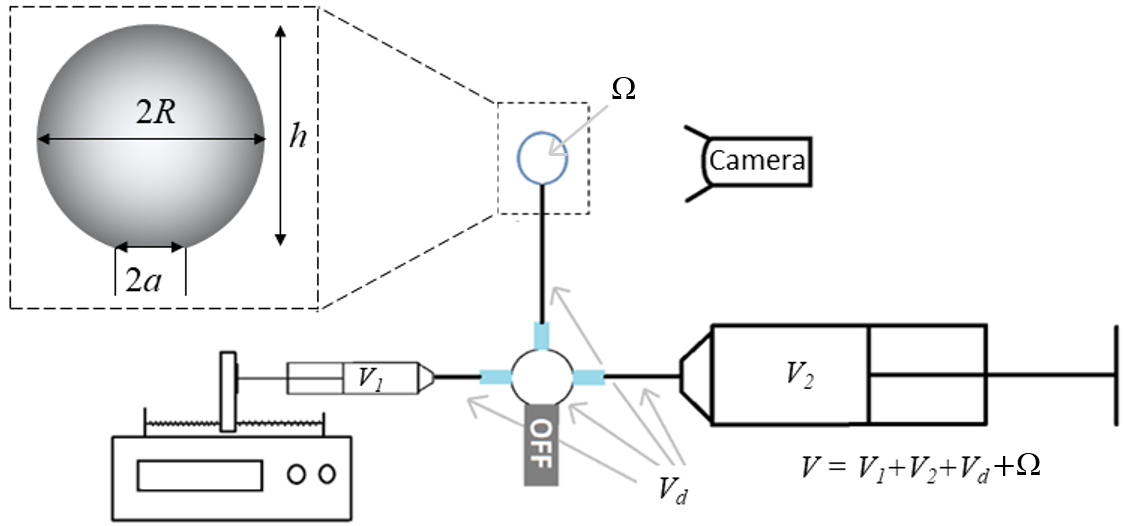}
    \caption{Experimental setup: the bubble of radius of curvature $R$ and height $h$ is connected to a large reservoir of volume $V$ constituted by the volume of the two syringes, $V_1$ and $V_2$, the dead volume of the connectors $V_d$ and the volume of the spherical cap, $\Omega$, of the bubble. The volume $V_1$ is decreased at a flow rate $Q$ thanks to a syringe pump. }
    \label{Fig:1setup}
\end{figure}

The soap film is made of a mixture of Sodium Dodecyl Sulfate (SDS) at a concentration of 24 mmol$/$L, which is 3 times larger than the CMC, 20\% of glycerine and deionized water. The solution is used at least 3 days after it has been made to ensure that the hydrolysis of SDS into dodecanol is achieved\cite{Bergeron1992}. The liquid/air surface tension, $\gamma$, is measured prior to any experiment using the pendant drop technique\cite{daerr2016pendentdrop} and we systematically found $\gamma=23 \pm 2$ mN/m. The soap film is deposed at the extremity of a needle - tube in the following - of external radius, $a$, ranging between $0.3$ and $0.83$ mm. The other extremity of the tube is connected to a reservoir composed of two syringes of volumes $V_1$ and $V_2$ (see Fig. \ref{Fig:1setup}). The total volume of the reservoir $V$ - which includes $V_1$, $V_2$, $V_d$ the dead volume of the valve and the tube and $\Omega$ the volume comprised between the film and the outlet of the tube - varies between $1$ and $50$ mL. A first syringe, connected to a syringe pump (KdScientific), is used to reduce the volume of the reservoir at a flow rate $-dV_1/dt=Q$, with $Q$ equal to $1$ or $2$ $\mu$L/s. A second syringe serves to change the initial total volume of the reservoir $V(t=0)=V_0$. The deformation of the soap film is monitored by a camera Marlin from Allied Vision.

At $t=0$, the syringe pump and the camera are triggered simultaneously (the error associated with this manual triggering is estimated at less than one second). The reduction of the volume of the reservoir increases the pressure and bends the soap film. The liquid film is much softer than the rest of the elements containing the compressed gas (syringe tube and connectors), then we assume that only the film is deformable.
To avoid premature rupture of the soap film, a transparent plastic box is placed around the bubble to limit its evaporation. This allows an easy observation of stable bubbles for several minutes. We made sure that this plastic box is not completely airtight so that the external pressure around the bubble is the atmospheric pressure.

\section{Experimental results}

\begin{figure}
    \centering
    \includegraphics[width=9cm]{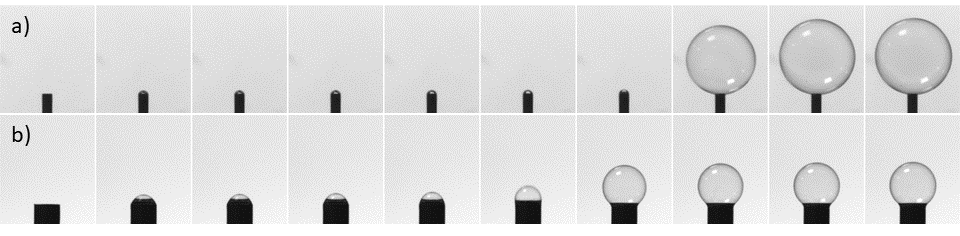}
    \caption{These two image sequences shows two bubbles swelled from reservoirs of identical volume $V_0=10$ mL. In a) $a=0.3$ mm and in b) $a=0.83$ mm. The thumbnails are separated by $5$ ms except for the first ones that display the tube before the beginning of the  soap films' compression. In a way to superposed the two swellings $Q= 2$ $\mu$L/s in a) and $Q= 1$ $\mu$L/s in b).}
    \label{Fig:2image_bulleformation}
\end{figure}

\begin{figure}
    \centering
    \includegraphics[height=6.5cm]{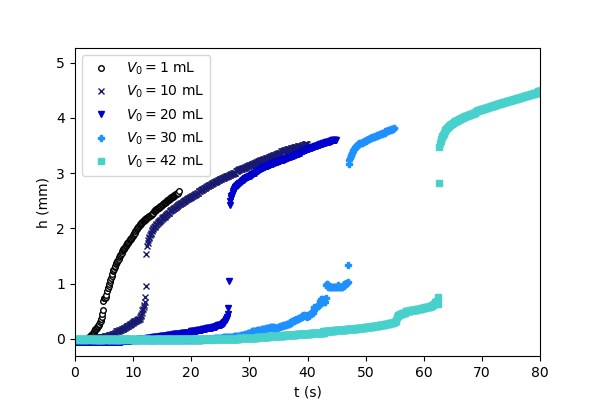}
    \caption{Bubble's height, $h$ as a function of time, $t$ for  $a=0.83$ mm, $Q$ = 2 $\mu$L$/$s and different $V_0$. For $V_0$ = 42, 30 and 20 mL, the bubbles inflate very rapidly while for $V_0$ = 1 and 10 mL, the bubbles inflate continuously. For each curve, the time origin is adjusted so that the data collapses at long times. The recording of the different curves is stopped just before the explosion of the bubbles.}
    \label{Fig:3hauteur_bulle}
\end{figure}

In our experiments, we have observed that it exist two clearly different regimes for a bubble to swell, a quasi-static one and a second highly dynamic. Those two regimes are illustrated in Fig. \ref{Fig:2image_bulleformation}.a) and b). In both cases $V_0$ is identical while $a$ is $2.8$ times bigger in b) than in a). The bubble in Fig. \ref{Fig:2image_bulleformation}.b), swelled from a large tube, continuously inflates step by step while the swelling of the bubble in Fig. \ref{Fig:2image_bulleformation}.a) is unstable and takes place in less than $5$ ms.
To go further, we report in Fig. \ref{Fig:3hauteur_bulle} the evolution of $h$, the height of the bubble, defined in Fig.\ref{Fig:1setup}, as a function of time, $t$, at same flow rate and radius but for different $V_0$. As we can see, we can make a distinction between two regimes, a first for $V_0 < 10$ mL where the growth of the bubble is continuous and a second for $V_0 \geq 10$ mL, where the swelling is unstable. In this second regime, the curves are S-shaped with a near-vertical zone meaning that the height of the bubble, $h$, changes from one to several millimeters in less than $5$ ms. 

The non-monotonic evolution of the radius of curvature, $R$, of the soap film is a crucial point to explain the distinction of regime observed in experiment. This radius is both constrained by $a$, the radius of the tube, and the evolution of the pressure, $P$, which follows the Laplace's law $P=P_0+4 \gamma/R$, where the factor 4 arises from the presence of two liquid/air interfaces. 

At first, the pressure in the reservoir is identical to the atmospheric pressure and the film is flat, thus $R \rightarrow \infty$. When the volume of the reservoir decreases due to the push syringe action, the pressure increases. Then, the soap film bends, $R$ decreases and Laplace-over pressure increase in agreement. However, due to geometrical constraint, $R$ cannot reach a value smaller than $a$ which correspond to a maximal pressure $P^*=P_0+4 \gamma/a$. At this point any compression of the volume by the push syringe triggers a second step of the film dynamic. The compression of the reservoir by the push syringe can not be balance with an increase of the pressure because $R$ is now increasing with the bubble bending. The bubble is in a non-equilibrium state in which compressed air has been store in the reservoir. To recover an equilibrium state, the bubble as to inflate until that the decompression of the stored air is done. This event has a very short characteristic time fully separate from the push syringe speed and an amplitude dependant of the quantity of compressed air stored. After, any further compression from the syringe pump is compensated by an increase of the bubble size.

\section{Model}

To understand these results, we write simple thermodynamic arguments stemming from the conservation of $n$, the number of gas moles in the reservoir. This is valid if the whole system is gas tight, hence if the rate of mole transfer $dn/dt$ due to the permeability $k$ of the soap film is negligible. From Fick's law, $dn/dt=-kA\Delta C$, where $k$ is the soap film permeability, $A\approx 4 \pi R^2$ the area of the bubble and $\Delta C$  the difference of gas concentration between the reservoir and the atmosphere surrounding the bubble. Using $R\sim $1 mm, $k \sim 1$ mm$/$s, a typical value from literature for SDS surfactants without salts \cite{Hadji2020,Krustev2002}, and $\Delta C=4 \gamma/(R R_u T_0)$, where $R_u=8.31$ J$/$mol$/$K is the universal gas constant and $T_0=298$ K the room temperature, we find $dn/dt \sim 2$ $10^{-10}$ mol$/$s. As the bubble typically forms in 100 s, the variation of moles in the bubble due to the permeability of the soap film is $\delta n \sim 2$ $10^{-8}$ mol. This is very small when compared to $n_0= P_0V_0/(R_u T_0)\approx 1.2 10^{-3}$ mol, thus we assume the system to be air tight and consider $n$ to be constant. Writing the conservation of $n$ for an isothermal transformation, yields:

\begin{equation}
    1=(1+\frac{4\gamma}{P_0R})(1+\frac{\Omega}{V_0} - \frac{Qt}{V_0})
    \label{equation_exact}
\end{equation}

Where $\Omega$ is the volume of the spherical cap above the tube of radius $a$. Since $R>a$, with $a$ ranging between 0.3 and 0.83 mm, $\frac{4\gamma}{RP_0}<\frac{4\gamma}{aP_0}<<1$, we make a Taylor expansion of Eq. \ref{equation_exact} and express the geometrical quantities $\Omega$ and $R$ as a function of $h$, the height of the spherical cap (see Fig. \ref{Fig:1setup}), using the geometrical relations $2hR=h^2+a^2$ and $\Omega=\pi h/2(a^2+\frac{h^2}{3})$. We also introduce the dimensionless parameters $x=h/a$, $\displaystyle \tau = \frac{2Qt}{\pi a^3}$, so that Eq. \ref{equation_exact} finally writes:
\begin{equation}\label{eq:3}
    \tau=x\left(1+\frac{x^2}{3}\right)+
    \frac{Bx}{x^2+1}
\end{equation}

\begin{figure}
    \centering
    \includegraphics[height=6.5cm]{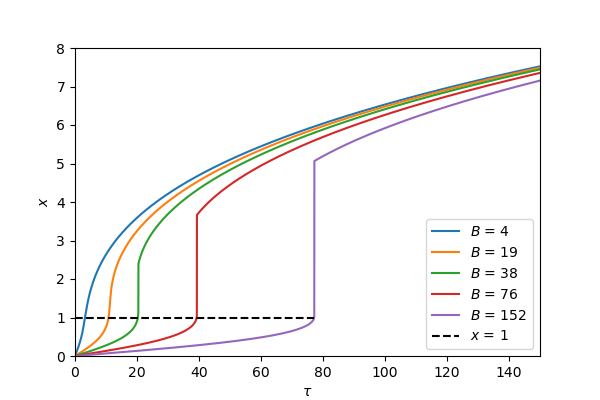}
    \caption{The theoretical dimensionless height $x$ as a function of dimensionless time $\tau$ for different values of $B$ calculated using Eq. \ref{eq:3}. Two regimes are observed: for $B\leq 19$, $x$ is defined unequivocally as a function of $\tau$ and the slope $dx/dt$ always reaches a finite value, whereas for $B\geq 38$, $x$ is multivalued and the slope $dx/dt$ reaches an infinite value.}
    \label{Fig:x_tau}
\end{figure}

With $\displaystyle B= \frac{16 \gamma V_0}{\pi a^4 P_0}$. The numerical resolution of Eq. \ref{eq:3} is plotted in figure \ref{Fig:x_tau} for various values of $B$. Two types of bubble growth are observed: for small values of $B$, $x$ increases as $\tau$ increases and the bubble formation is monotonic and proceeds continuously. This is in agreement with the observations of Fig. \ref{Fig:2image_bulleformation} and \ref{Fig:3hauteur_bulle}, which revealed continuous bubble formation for large values of $a$ and small values of $V_0$. For larger values of $B$ (typically $B\geq38$ on figure \ref{Fig:x_tau}), the curves corresponding to the numerical solution of Eq. \ref{eq:3} are S-shaped with non-monotonic variation of $\tau$ as a function of $x$, which is not physical since $\tau$ - the dimensionless time - should always increases. Thus, when $d x/d\tau$ is negative, there is no physical solution to the equation, and the dimensionless height suddenly jumps from one value to another. In the following, we call $x_1$ the maximum value of $x$ before the jump and $x_2$ the minimum value of $x$ after the jump.

To determine $x_1$ and $x_2$, we proceed as follows. We first calculate $x_1$, the dimensionless height at the onset of the formation of an unstable bubble, for which $\left. \frac{d\tau}{dx}\right|_{x=x_1}=0$. Hence $x_1$ is a solution of:
 \begin{equation}\label{eq:condition_stabilite}
      B^{-1}\left(1+x^2\right)^3-x^2+1=0
 \end{equation}
Using $X=1+x^2$, Eq. \ref{eq:condition_stabilite} can be reduced to a polynomial of degree 3:
\begin{equation} \label{eq:polynome}
    B^{-1} X^3-X+2=0 
\end{equation}
We seek for solutions larger than one using the Cardan method \cite{Cardan1545}. For $B \leq 27$, there is no real solution larger than one for Eq. \ref{eq:polynome}, hence $\displaystyle \frac{d\tau}{dx}>0$ and the formation of the bubble is continuous. 

For $B\geq 27 $, two real solutions exist, but only one, $X_1$, is larger than one, increases with $B$, and is physically consistent when $B \rightarrow \infty$: 
\begin{equation}
    X_1=2\sqrt{\frac{B}{3}}\cos{\left[\frac{1}{3}\arccos{\left(-3\sqrt{\frac{3}{B}}\right)+\frac{4\pi}{3}}\right]}\label{Eq:X1}
\end{equation}

Thus, the instability is triggered, as soon as $x>x_1$, where $x_1=\sqrt{X_1-1}$, with $X_1$ given by Eq. \ref{Eq:X1}.

To determine $x_2$, the dimensionless height of the bubble after the unstable swelling, we assume that the swelling is instantaneous and writes $\tau(x_1)=\tau(x_2)$ using Eq. \ref{eq:3}. This leads to a 4th order polynomial equation in $x_2$:
\begin{equation}=
    x_2^4+x_1x_2^3+(x_1^2+4)x_2^2+x_1\left(1-\frac{3B}{x_1^2+1}\right)x_2+3+x_1^2+\frac{3B}{x^2+1}=0
\label{eq:polynome_4}
\end{equation}    

\begin{figure}
    \centering
    \includegraphics[width=8cm]{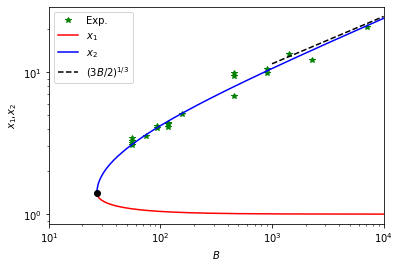}
    \caption{Analytical solutions $x_1$ of Eq. \ref{Eq:X1} and $x_2$ of Eq. \ref{Eq:x2_polynome_deg3} as a function of $B$. The two curves meet for $B=27$ at $x_1=x_2=\sqrt{2}$. For $B<27$, $x_1$ and $x_2$ are not defined as the bubble inflates continuously. For $B\xrightarrow{} \infty $, $x_1 \xrightarrow{} 1$ and $x_2 \sim (\frac{3}{2}B)^{1/3}$. The green stars correspond to experimental data.}
    \label{Fig:x1_x2_function_B}
\end{figure}

Since $x_1$ is also a solution of Eq.\ref{eq:polynome_4} we factor by $(x_2-x_1)$ to reduce the polynomial of degree 4 to a polynomial of degree 3: 

\begin{equation}
    x_2^3+2x_1 x_2^2 + (4+3{x_1}^2)x_2 - \left(\frac{3B}{ x_1(1+{x_1}^2)} +\frac{3}{x_1}+x_1\right)=0 \label{Eq:x2_polynome_deg3}
\end{equation}
Eq. \ref{Eq:x2_polynome_deg3} can also be solved analytically using the Cardan method after the change of variable $X_2=x_2+\frac{2x_1}{3}$ \cite{Cardan1545}, which leads to:
\begin{equation}
    X_2= \sqrt[3]{\frac{1}{2}(-q+\sqrt{\frac{D}{27}})}-\sqrt[3]{\frac{1}{2}(q+\sqrt{\frac{D}{2}})} \label{Eq:X2}
\end{equation}
with $D = 27q^2+4p^3$, $q=\frac{43}{27}x_1^3+\frac{63}{27}x_1-\frac{3Bx_1}{(1+x_1^2)}$ and $p=\frac{5}{3}x_1^2+4$. $x_1$ and $x_2$ are plotted in Fig. \ref{Fig:x1_x2_function_B} and discussed in the following section.

\section{Results}

We now discuss the outcomes of the model. In Figure \ref{Fig:x1_x2_function_B}, we plot the simple analytical expressions of $x_1$ and $x_2$ as a function of $B$.
For $B<27$, $x_1$ and $x_2$ are not defined and the bubble growth proceeds continuously. For $B=27$, Eq. \ref{Eq:X1} and \ref{Eq:X2} insures $X_1=3$ and $X_2=\frac{5}{3}\sqrt{2}$, hence $x_1=x_2=\sqrt{2}$, as highlighted by the black dot of coordinate $(27,\sqrt{2})$, which superimposes with the two curves of Fig. \ref{Fig:x1_x2_function_B}. For $B\xrightarrow{}\infty$, Eq. \ref{eq:condition_stabilite} gives $x_1 \xrightarrow{}1$ as observed in Fig. \ref{Fig:x1_x2_function_B}. Since $B$ increases with $V_0$ and decreases with $a$, this  suggests that for large $V_0$ and$/$or small $a$, the instability is triggered as soon as $h\xrightarrow{}a$, hence when the bubble reaches a shape very close to the hemisphere in agreement with the Laplace pressure limit set by the radius of the tube. Then imposing $x_1$ in Eq. \ref{Eq:x2_polynome_deg3} yields to $x_2 \sim \left(\frac{3}{2}B\right)^{1/3}$. In this asymptotic limit, which correctly reproduces the full calculation of $x_2$ for $B \geq 10^3$ as highlighted by the dashed line Fig. \ref{Fig:x1_x2_function_B}, the height of the bubble at the end of the instability is of the order of $a\left(\frac{3}{2}B\right)^{1/3}$. In this limit of large $B$, the bubble is quasi-spherical and its volume $\Omega$ right after the jump is $\Omega=\frac{\pi}{6}h^3$. Therefore, our model predicts a final bubble volume equal to $V_0\frac{4\gamma}{a P_0}$, which is surprisingly proportional to $V_0$ the volume of compressed gas upstream of the constricted zone modulated by the ratio of the Laplace pressure over the atmospheric pressure.

The comparison of this theoretical findings with the experimental data is not immediate for the following reasons. First, the determination of $ax_1$, the height of the bubble at the onset of the instability is delicate due to the small range of variation of $x_1$, so we do not propose experimental data points on Fig. \ref{Fig:x1_x2_function_B} conserning $x_1$.Second, the data of figure \ref{Fig:3hauteur_bulle} shows a very fast growth of the bubble, but not instantaneous. Thus, to experimentally extract the height $ax_2$ corresponding to the end of the instability, we use the following arbitrary criterion: the bubble is in the unstable mode as soon as $dh/dt$ is greater than $af$ where $f$ is the acquisition frequency of the camera. Despite this arbitrary criterion, the corresponding experimental data show a remarkable agreement with the model as illustrated in figure \ref{Fig:x1_x2_function_B}. The model thus confirms the importance of $V_0$ and $a$ for bubble sizing and gives a direct relation between those parameters which could have direct application in microfluidic engineering processes.  

\section{Discussion}
The system we describe - namely ejection of a large volume of gas when the pressure in the microfluidic reservoir exceeds a critical value - is analogous to what could be observed when following the volume of gas ejected from a macroscopic pressure cooker equipped with a weighted valve. In these cookers, the charging phase where the pressure increases in the tank is contained by the weight of the valve, is followed by a discharging phase, where a large volume of gas is ejected very quickly when the pressure exceeds the threshold supported by the valve. Beyond that, the gas flow rate out of the cooker remains constant. In the problem we study, the constriction of the tube, $a$, that imposes the maximum capillary pressure that the system can support is then equivalent to the valve of the pressure cooker. Recently, Keiser \& al \cite{Keiser2022} have shown that a similar behavior can also be observed in a dead-end  microchannel containing a constriction, initially filled with water. The unstability is then driven by the pervaporation of the liquid through the channels. Yet, in their case, the water being incompressible, it is the compliance of the elastic channels that allows the variation of pressure of the water. As for our system, the kinetics of fluid escape depends on the volume under tension. Magdelaine \& al.\cite{Magdelaine2019} who studied a gaseous system very similar to the one considered here where the volume of compressed gas ejects into water rather than into a bubble, also highlights the importance of the volume of the pressurized reservoir. By adopting a very different formalism from ours and introducing the pinching kinetics of the gas jet ejected into the water, they produce a comprehensive model predicting the number of bubbles formed during the compression of a gaseous syringe. 
In general, in these two-phase systems, it is the compliance of the system, whether it comes from the compressibility of the gas or the elasticity of the microfluidic channels, which is at the origin of this instability as thoroughly discussed in \cite{vanLoo2016} for two-phase microfluidics flow.

In view of these results, two points seem interesting to discuss.

The first point of interest concerns the formation of monodisperse bubbles in microfluidic geometries where the interface is confined in a constriction, like flow-focusing. Experimentalists in this field are well aware that the bubble size distributions produced in this type of geometry when the gas phase is flow-rate driven are more difficult to control and less peaked (with standard deviations higher than 20$\%$) than when the gas is driven at controlled pressure \cite{Ward2005}. This explains why pressure-driven gas control is often preferred to flow-rate gas control. Our work sheds light on this point : taking typical values $a=100$ $\mu$m, $V_0=1$ mL, it comes $B\sim 1000$, which clearly shows that those devices are in the unstable regime highlighted here. This suggests that the volume of the syringe containing the gas, $V_0$, a parameter usually not considered, must be taken into account to set the bubbles size.

Second, the proposed model, in very good agreement with the experiments, allows to predict the unstable growth regime ($B>27$) as well as the amplitude of this phase, set by $a(x_2-x_1$). Since it is based on quasi-static arguments, it does not perfectly capture the growth dynamics of the bubble in the unstable regime. Indeed, for $B>27$, we predict that the height $h/a$ jumps from $x_1$ to $x_2$ instantaneously (see Fig. \ref{Fig:x_tau}), which is neither physical nor confirmed by experiments. As can be seen in Fig. \ref{Fig:3hauteur_bulle}, the growth is very fast but not infinite because in practice, this expansion regime is limited by a dissipative process being either inertia of the gas, viscosity of the gas or liquid or rheology of the interface. A detailed follow-up of the bubble growth kinetics using a high-speed camera in the limit where it is limited by the interfacial rheology, seems a promising prospect for this work, since it could open the way to a new characterization of the interfacial rheology of surfactants in a nearly spherical geometry. Indeed, it has been recently shown that "Capillary pressure Elastometry", which consists in analyzing the quasi-static pressure-deformation curves for a bubble in this type of geometry, allows to analyse the elastic properties of interfaces \cite{Ginot2021}. The simple model proposed here should allow to extend the field of application of "Capillary pressure Elastometry" to the dynamic response of interfaces.

\section{Conclusion}
We have shown that the growth of bubbles blown in geometries with non-zero compliance can exhibit unstable regimes. Our experimental and theoretical study reveals the importance of the coupling between the constriction zone on which the interface is anchored and the volume of the tank in which the gas is compressed. The use of these unstable regimes to probe elongational interfacial rheology seems to us among the most promising perspective of this work. 

\section*{Author Contributions}
M.G. designed the research, conducted the experiments and made the model. M.G and E.L discussed the results and wrote the article.

\section*{Conflicts of interest}
There are no conflict to declare.

\begin{acknowledgments}
The authors thank Jerome Giraud for his technical help and Ludovic Keiser, Philippe Marmottant and Benjamin Dollet for fruitful discussions. This work is supported by funding from Pack Ambition Recherche 2021 of the region AURA - SELFI project.
\end{acknowledgments}

\bibliography{apssamp.bib}

\end{document}